\documentclass[journal=jacsat,manuscript=article]{achemso}

\usepackage{chemformula} 
\usepackage[T1]{fontenc} 

\usepackage{graphicx} 
\usepackage[version=4]{mhchem} 
\usepackage[]{siunitx}

\author{Marijn Rikers}
\affiliation[Friedrich Schiller University Jena]{Institute of Solid State Physics, Friedrich Schiller University Jena, Max-Wien Platz 1, 07743, Jena, Germany}
\alsoaffiliation[Abbe Center of Photonics]{Institute of Applied Physics, Abbe center of Photonics, Friedrich Schiller University Jena, Albert-Einstein-Straße 6, 07745, Jena, Germany}
\alsoaffiliation[Australian National University]{Department of Quantum Science and Technology, Research School of Physics, Australian National University, 60 Mills Rd., Canberra, ACT 2601, Australia}
\email{marijn.rikers@uni-jena.de}

\author{Ayesheh Bashiri}
\affiliation[Friedrich Schiller University Jena]{Institute of Solid State Physics, Friedrich Schiller University Jena, Max-Wien Platz 1, 07743, Jena, Germany}
\alsoaffiliation[Abbe Center of Photonics]{Institute of Applied Physics, Abbe center of Photonics, Friedrich Schiller University Jena, Albert-Einstein-Straße 6, 07745, Jena, Germany}

\author{Aleksandr Vaskin}
\affiliation[Friedrich Schiller University Jena]{Institute of Solid State Physics, Friedrich Schiller University Jena, Max-Wien Platz 1, 07743, Jena, Germany}

\author{Ángela Barreda}
\affiliation[Friedrich Schiller University Jena]{Institute of Solid State Physics, Friedrich Schiller University Jena, Max-Wien Platz 1, 07743, Jena, Germany}
\alsoaffiliation[University Carlos III]{Department of Electronic Technology, University Carlos III of Madrid, Avda. de la Universidad 30, 28911 Leganés, Spain}

\author{Duk-Yong Choi}
\affiliation[Australian National University]{Department of Quantum Science and Technology, Research School of Physics, Australian National University, 60 Mills Rd., Canberra, ACT 2601, Australia}

\author{Michael Steinert}
\affiliation[Abbe Center of Photonics]{Institute of Applied Physics, Abbe center of Photonics, Friedrich Schiller University Jena, Albert-Einstein-Straße 6, 07745, Jena, Germany}

\author{Thomas Pertsch}
\affiliation[Abbe Center of Photonics]{Institute of Applied Physics, Abbe center of Photonics, Friedrich Schiller University Jena, Albert-Einstein-Straße 6, 07745, Jena, Germany}
\alsoaffiliation[Fraunhofer IOF]{Fraunhofer-Institute for Applied Optics and Precision Engineering IOF, Albert-Einstein-Str. 7, 07745 Jena, Germany}
\alsoaffiliation[Max Planck School of Photonics]{Max Planck School of Photonics, Hans-Knöll-Str. 1, 07745 Jena, Germany}

\author{Isabelle Staude}
\affiliation[Friedrich Schiller University Jena]{Institute of Solid State Physics, Friedrich Schiller University Jena, Max-Wien Platz 1, 07743, Jena, Germany}
\alsoaffiliation[Abbe Center of Photonics]{Institute of Applied Physics, Abbe center of Photonics, Friedrich Schiller University Jena, Albert-Einstein-Straße 6, 07745, Jena, Germany}
\alsoaffiliation[Max Planck School of Photonics]{Max Planck School of Photonics, Hans-Knöll-Str. 1, 07745 Jena, Germany}

\title{Polarization Dependent Enhancement of Magnetic Dipolar Emission with Silicon Nanodimers}

\abbreviations{}

\begin{document}

\begin{abstract}
\ce{Eu(TTA)3} complexes are used as an emission source in the presence of high refractive index dielectric nanostructures. These nanostructures support Mie-type resonances that modify the local density of optical states. Specifically, the silicon dimer provides polarization-dependent electric and magnetic field enhancement in the dimer gap to modify the electric dipolar and magnetic dipolar emissions of the \ce{Eu^{3+}} at \SI{610}{ \nano\meter} and \SI{590}{\nano\meter}, respectively. Finite element method simulations are used to determine the optimal parameters for the sample and to demonstrate the polarization-dependent emission enhancement of dipolar emitters in the gap. A two-step electron beam lithography process is used to fabricate the hybrid nanoscopic structures, with a \ce{Eu3+} doped electron beam resist located only in the center of the dimer. The results demonstrate the potential of these nanostructures to selectively tailor the emission of the two distinct dipolar transitions by engineering the resonant nanostructures. Our work highlights the potential of magnetic light-matter interactions as a novel degree of freedom.
\end{abstract}

\section{Introduction}
\begin{figure}[ht]
	\centering
	\includegraphics[width = 0.9\textwidth]{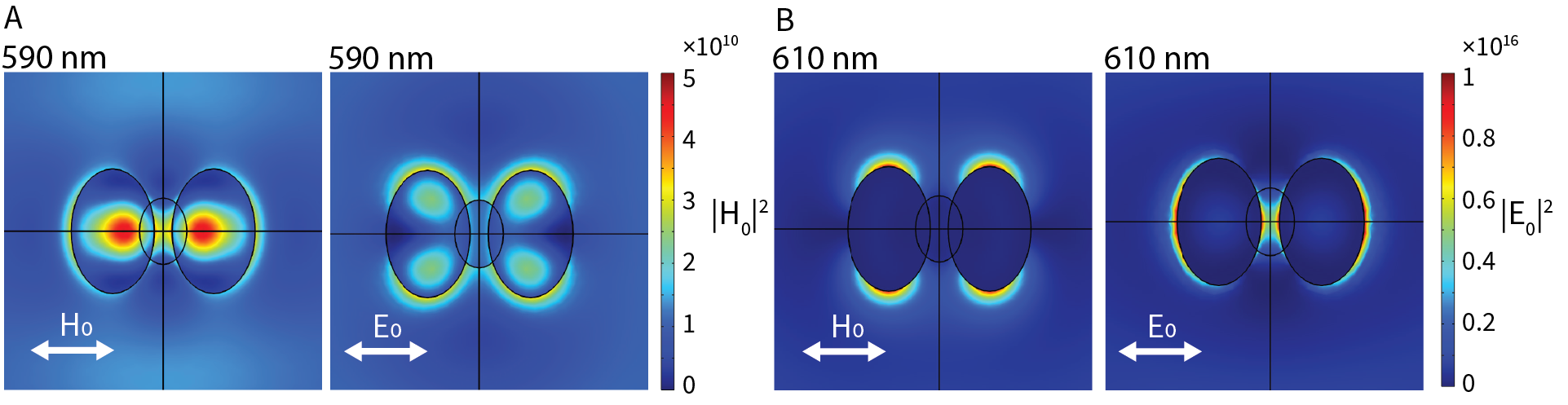}
	\caption{a) Near-field maps of silicon dimers with optimal geometry for magnetic field intensity enhancement at $\lambda = \SI{590}{ \nano\meter} $, where the incident radiation is linearly polarized with the electric (right figure) or magnetic (left figure) field across the gap. b) Near-field maps of the silicon dimer for the electric field intensity enhancement at $\lambda = \SI{610}{ \nano\meter} $, where the incident radiation is linearly polarized with the electric (right figure) or magnetic (left figure) field across the gap.}
	\label{fig:Simmulation}
\end{figure}

The decay rate of quantum emitters can be modified by changing the environment where the emitter is located using the Purcell effect \cite{purcellSpontaneousEmissionProbabilities1946}. The Mie-type resonances that high refractive index dielectric nanostructures support can provide a higher local density of optical states at which an emitter can be coupled in comparison to a background medium, enhancing its emission. In addition, the interferential effects between electric and magnetic modes allow for governing the directionality of the emitted radiation \cite{vaskinLightemittingMetasurfaces2019}.
While the magnetic component of light is generally much weaker than the electric component, trivalent Lanthanide ions like Eu3+ have magnetic dipolar (MD) \ce{{}^5D0\bond{->}{}^7F1} and electric dipolar (ED) \ce{{}^5D0\bond{->}{}^7F_i} $i\in\{0,2,4,6\}$ transitions in the VIS range that are comparable in magnitude \cite{ofeltIntensitiesCrystalSpectra1962a, blasseEu3FluorescenceMixed1966}. Moreover, magnetic and electric dipolar transitions exhibit different responses to their respective LDOS \cite{noginovaMagneticDipoleBased2008}. In this work, we demonstrate the polarization-dependent emission enhancement, with a silicon dimer, of the \ce{Eu(TTA)3} complex. 
Our previous work \cite{rikersFabricationCharacterizationNanoscopic2022} has shown the metal-organic complex \ce{Eu(TTA)3} is soluble in the negative tone electron beam lithography (EBL) resist \ce{ma\bond{-}N}, allowing for the precise fabrication of a nanoscopic emitter with a resolution of approximately \SI{100}{ \nano\meter} that remains fluorescent after exposure. This enables the deterministically placed \ce{Eu3+} emitters exclusively in the dimer's gap.
The magnetic component of light provides an additional degree of freedom, enabling polarization routing. The design of a silicon dimer is presented with a polarization-dependent response that results in the separate routing of magnetic dipolar (MD) and electric dipolar (ED) emissions of the \ce{Eu^{3+}} ion at \SI{590}{ \nano\meter}  and \SI{610}{ \nano\meter} , respectively, into distinct polarizations. This work emphasizes the potential of magnetic light-matter interactions as a new means of light emission control. 
Dielectric nanoresonators can support both electric and magnetic dipolar and higher multipole order resonances \cite{staudeTailoringDirectionalScattering2013}, which can be combined in a dimer structure to support complex electromagnetic responses. When the induced magnetic or electric dipoles of the nanoresonators are aligned across the gap, a magnetic or electric hotspot, respectively, is observed \cite{bakkerMagneticElectricHotspots2015}. In this way, the dimers can enhance the decay rate of the electric and magnetic dipole transitions of an emitter located in the gap \cite{barredaMetalDielectricHybrid2021}. In particular, in our work, the dimer is designed to show an electric/magnetic $(\SI{610}{ \nano\meter} /\SI{590}{ \nano\meter} )$ dipole hot spot at the wavelength of the electric/magnetic dipole transition of \ce{Eu3+} emitters.
The optimal parameters for the dimer sample include a height of \SI{70}{ \nano\meter} , a short axis diameter of \SI{130}{ \nano\meter} , a long axis diameter of $1.5\times$ the short axis, and a gap width of \SI{30}{ \nano\meter}  which were determined using COMSOL Multiphysics. The near-field maps in Fig. \ref{fig:Simmulation} illustrate the polarization-dependent emission enhancement for magnetic or electric dipoles in the gap.

\section{Fabrication}
\begin{figure}[ht]
	\centering
	\includegraphics[width = 0.9\textwidth]{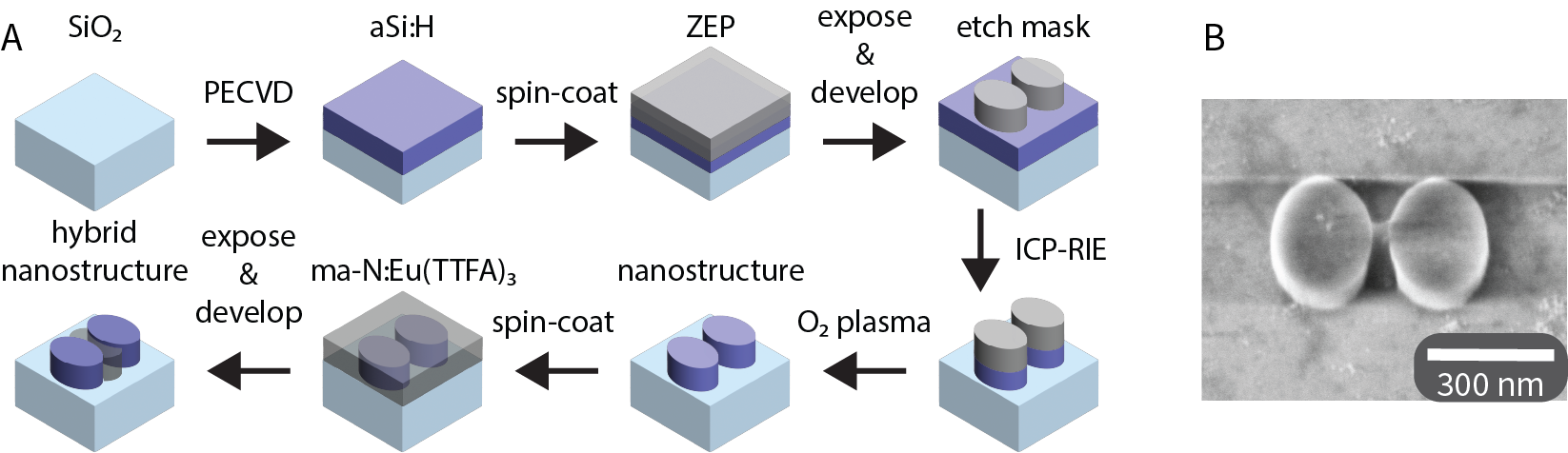}
	\caption{a) Schematic overview of the fabrication steps used to create the hybrid nanoscopic. b) scanning electron microscope image of the fabricated dimer \ce{$\alpha$Si$:$H} with the \ce{ma\bond{-}N$:$Eu(TTA)3} resist in the center.}
	\label{fig:Fabrication}
\end{figure}

To fabricate a hybrid nanostructure, a two-step electron beam lithography (EBL) process is employed, using alignment markers as reference points to align both exposures. The first step involves fabricating dimers from amorphous silicon (\ce{$\alpha$Si$:$H}). For the second step, europium-doped electron beam resist is exposed only in the center of the dimer, resulting in the hybrid nanostructure.
\ce{PECVD} is utilized to deposit a \SI{70}{\nano\meter} layer of \ce{$\alpha$Si$:$H} onto a \ce{SiO2} substrate. Next, a layer of \ce{ZEP 502} EBL resist is spin-coated onto the sample. The resist is exposed with a dose of \SI{10}{\milli\coulomb\per\square\centi\meter}. \ce{ZEP}, at high doses, functions as a negative-tone resist with a resolution of \SI{10}{ \nano\meter}  similar to \ce{PMMA} \cite{hooleNegativePMMAHighresolution1997}. The exposed \ce{ZEP} acts as the etching mask, and inductively coupled plasma reactive ion etching (ICP-RIE), is used to etch the \ce{$\alpha$Si$:$H} into the nanodimer.
To confine the \ce{Eu(TTA)3} emitters only in the gap, a previously reported process is utilized \cite{rikersFabricationCharacterizationNanoscopic2022}. \ce{Eu(TTA)3} doped \ce{Ma\bond{-}N 2401} (\ce{Ma\bond{-}N$:$Eu(TTA)3}) resist is spin-coated onto the sample. The alignment feature of the EBL system is utilized for the exposure, aligning the dot exposure to the center of the dimer gap. Subsequently, the center of the dimer is exposed with a single dot $(\text{dose} = \SI{64}{\femto\coulomb})$, and after developing the sample, the \ce{Ma-N$:$Eu(TTA)3} is only present in the center of the dimer. These steps are schematically represented in Fig. \ref{fig:Fabrication}a. An SEM image of the finished dimer structure is shown in Fig. \ref{fig:Fabrication}b.

\section{Conclusion}
The present work proposes a nanoscopic dimer for enhancing the polarization-dependent emission of \ce{Eu3+} for two independent wavelengths of \SI{590}{\nano\meter} and \SI{590}{\nano\meter} corresponding to the magnetic and electric dipole transitions, respectively. The magnetic dipolar transition \ce{{}^5D0\bond{->}{}^7F1} is enhanced at $\lambda=\SI{590}{\nano\meter}$ when the magnetic field is aligned across the gap of the dimer, while the electric dipolar transition \ce{{}^5D0\bond{->}{}^7F2} is enhanced at  $\lambda=\SI{610}{\nano\meter}$ when the electric field is aligned across the gap, for the \ce{Eu(TTA)3} complexes. The suggested nanoscopic photonic device can route separate wavelengths into differing polarization channels using the magnetic component of light.

\section{Acknowledgment}
This work is funded by Deutsche Forschungsgemeinschaft (DFG, German Research Foundation) through the International Research Training Group (IRTG) 2675 “Meta-ACTIVE”, project number 437527638. Furthermore, we thank the ANFF for providing the facilities to fabricate the sample.

\bibliography{ReferencesMetamaterials.bib}

\end{document}